\pdfoutput=1
\documentclass[aps,prc,reprint,superscriptaddress,showpacs,amsmath,amssymb,longbibliography,lengthcheck]{revtex4-1}
\usepackage[]{graphicx}
\usepackage{mathrsfs}
\usepackage{txfonts}

\begin{document}
 \title{The character and 
prevalence of third minima in actinide fission barriers} 
\author{Takatoshi Ichikawa}%
\affiliation{Yukawa Institute for Theoretical Physics, Kyoto University,
Kyoto 606-8502, Japan}
\author{Peter M{\"{o}}ller}
\author{Arnold J Sierk}
\affiliation{Theoretical Division, Los Alamos National Laboratory, Los
Alamos, New Mexico 87545, USA}
\date{\today}

\begin{abstract}
The double-humped structure of many actinide fission barriers is well
established both experimentally and theoretically.  There is also
evidence, both experimental and theoretical, that some actinide nuclei
have barriers with a third minimum, outside the second,
fission-isomeric minimum.  We perform a large-scale, systematic
calculation of actinide fission barriers to identify which actinide
nuclei exhibit third minima.
We find that only a relatively few nuclei accessible to experiment
exhibit third minima in their barriers, approximately nuclei
with proton number $Z$ in the range $88 \leq Z \leq 94$ and nucleon
number $A$ in the range $230 \leq A \leq 236 $. We find that the third
minimum is less than 1 MeV deep for light Th and U isotopes.
This is consistent with some previous experimental and theoretical results, but differs from some others.
We discuss possible origins of these incompatible results and  what
are the most realistic predictions of where 
third minima are observable.
\\[-0.25in] 

\end{abstract}
\maketitle

Third minima were first obtained in calculated fission potential-energy
surfaces in 1972 \cite{moller72:a}, in macroscopic-microscopic
calculations based on the modified-oscillator (Nilsson)
single-particle potential. A year later similar results were found
in macroscopic-microscopic calculations based on a substantially
different single-particle potential, namely the folded-Yukawa
potential \cite{moller74:a,moller74:b}, for two
different shape parameterizations.  Subsequently, other studies
found similar results \cite{bengtsson87:b}. Such studies have usually
found only a few actinide nuclei with third minima, mainly
elements in the range $88\leq Z \leq 94$ for isotopes in the range $
228 \leq A \leq 236$.  Normally, these minima are about 5 MeV above the
ground states and are surrounded by saddles rising to about one MeV
above the minimum.  In our discussion, we adopt the common
notation of $E_{\rm I}$, $E_{\rm II}$, and $E_{\rm III}$ for the
energies of the ground state minimum, the fission isomeric (second)
minimum and the third minimum, respectively, and
$E_{\rm A}$, $E_{\rm B}$, and $E_{\rm C}$ for the energies of the
three barrier peaks, starting from the inner one.

Since third minima appeared relatively consistently in
macroscopic-microscopic calculations with different single-particle
potentials and their results therefore appeared quite robust, it was
suggested by Ray Nix in 1973 \cite{moller74:a} that experimental
signatures of these third minima might be observed.  He also suggested
that the discrepancy between the calculated height of the inner peak
in the barrier, about 4 MeV, of light actinides and the consistent
experimental result of 6 MeV might be resolved if the experimental
barrier peak listed as 6 MeV high was actually $E_{\rm B}$ and not
$E_{\rm A}$.

Inspired by these suggestions, during the next decade Blons and 
\begin{figure*}[t] 
 \begin{center} 
 \includegraphics[width=0.75\linewidth]{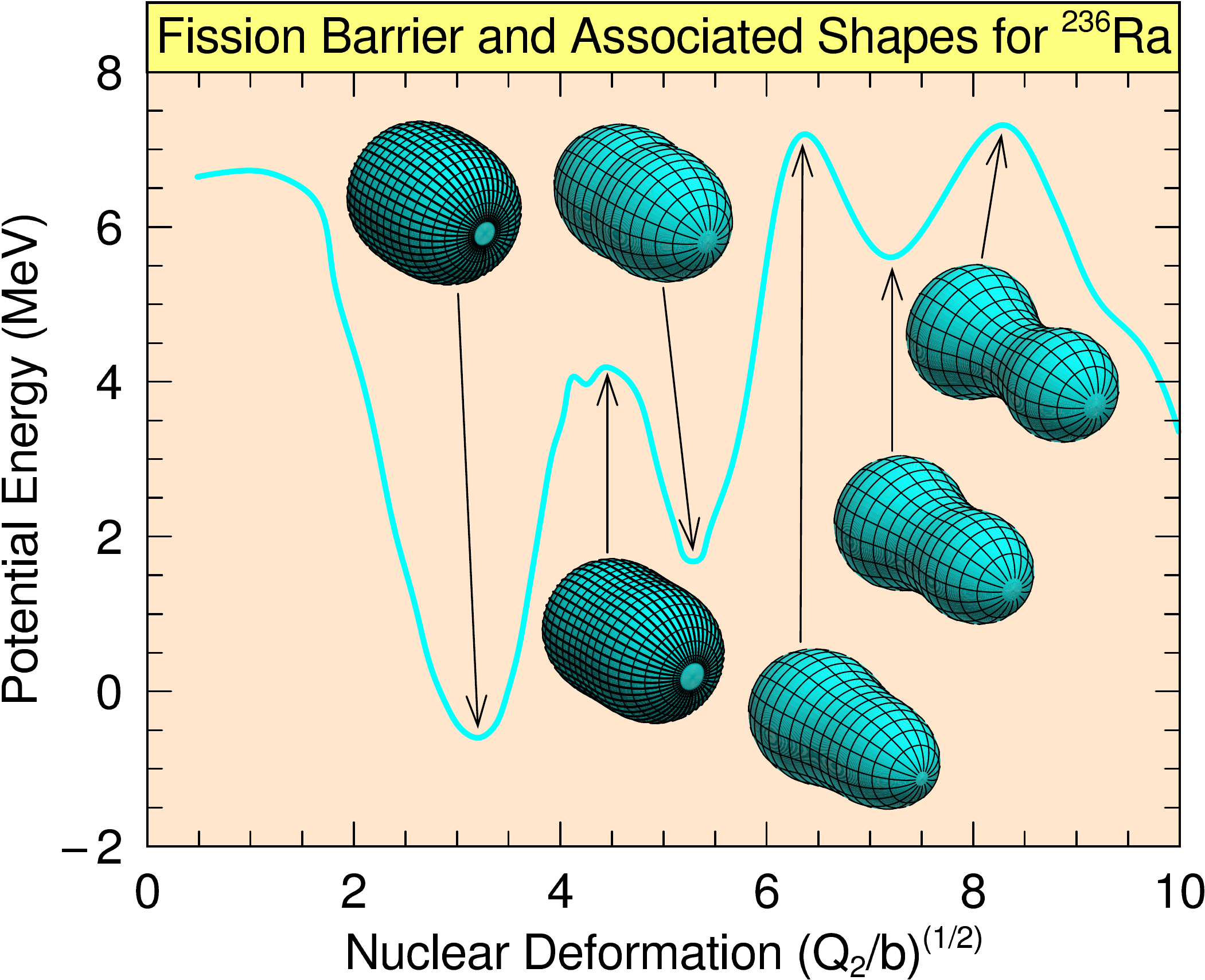} 
   \caption{Calculated fission barrier for $^{236}$Ra. This is one of the 
deepest third minima we find, for which the surrounding saddles are of
about equal height. The saddles rise about 1.5 MeV above 
\vspace{-0.3in} the third minimum.}
\label{barra}
 \end{center}
\end{figure*}
collaborators
carried out a series of increasingly detailed experiments aimed at
elucidating the properties of the third minimum
 \cite{blons75:a,blons78:a,blons88:a}. These
revealed rotational-like level structures at 5--6 MeV excitation energy,
\begin{figure*}[t] 
 \begin{center} 
 \includegraphics[width=0.78\linewidth]{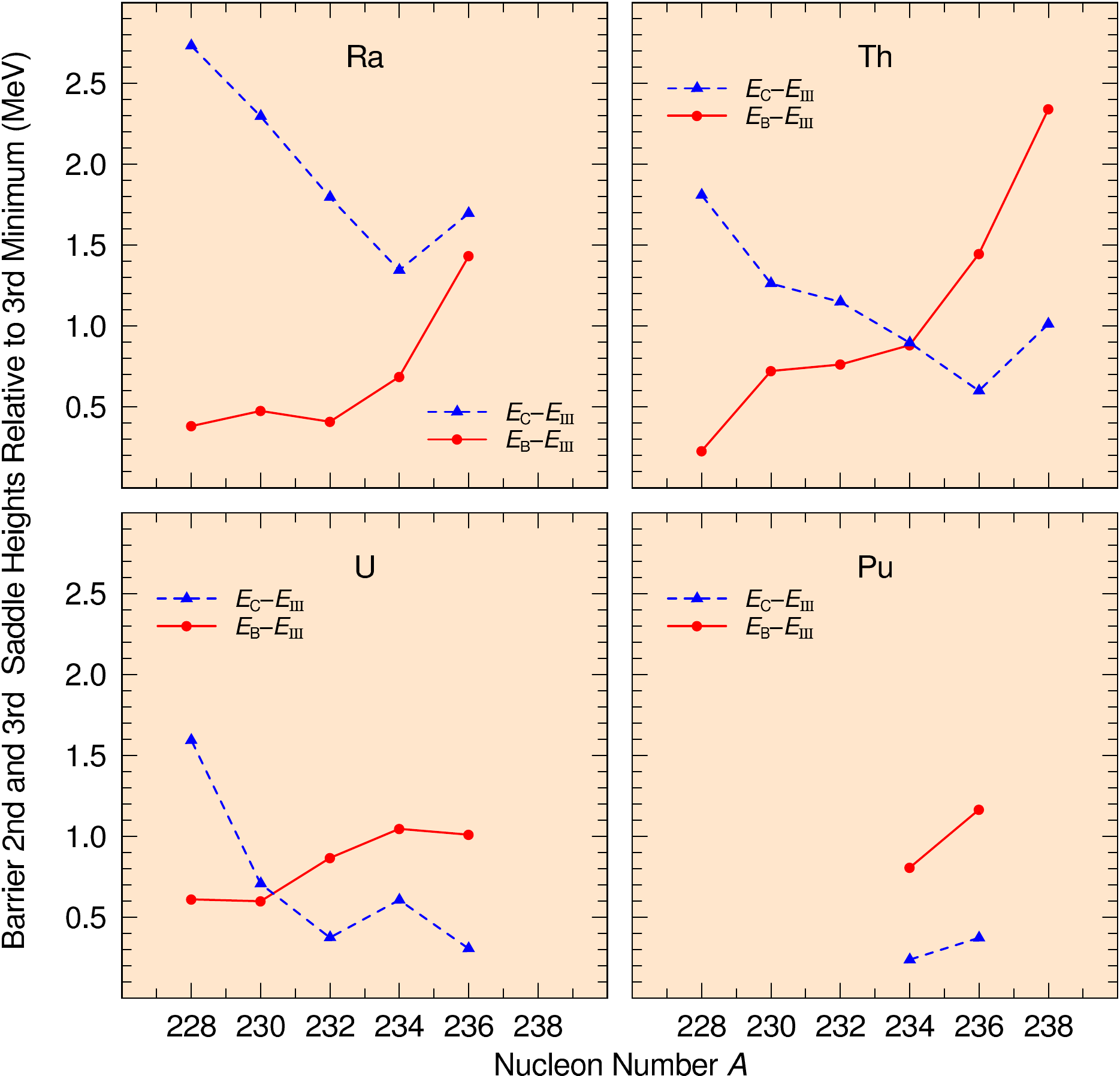} 
   \caption{Depth of the third minimum relative to surrounding
barrier saddle points for four heavy elements. In uranium,
$^{224}$U and $^{226}$U also exhibit third minima with $E_{\rm B}$
about 0.6 MeV and $E_{\rm C}$ about 2 MeV above the minimum for both these isotopes. U and Pu isotopes do not have third minima for systems heavier
than those plotted. Th and Ra do have 
third minima for systems heavier than those shown, but since we mainly
focus on relatively accessible nuclei here, we do not discuss \vspace{-0.3in}them.}
\label{thirdmin}
 \end{center}
\end{figure*}
\begin{figure}[t] 
 \begin{center} 
 \includegraphics[width=1.0\linewidth]{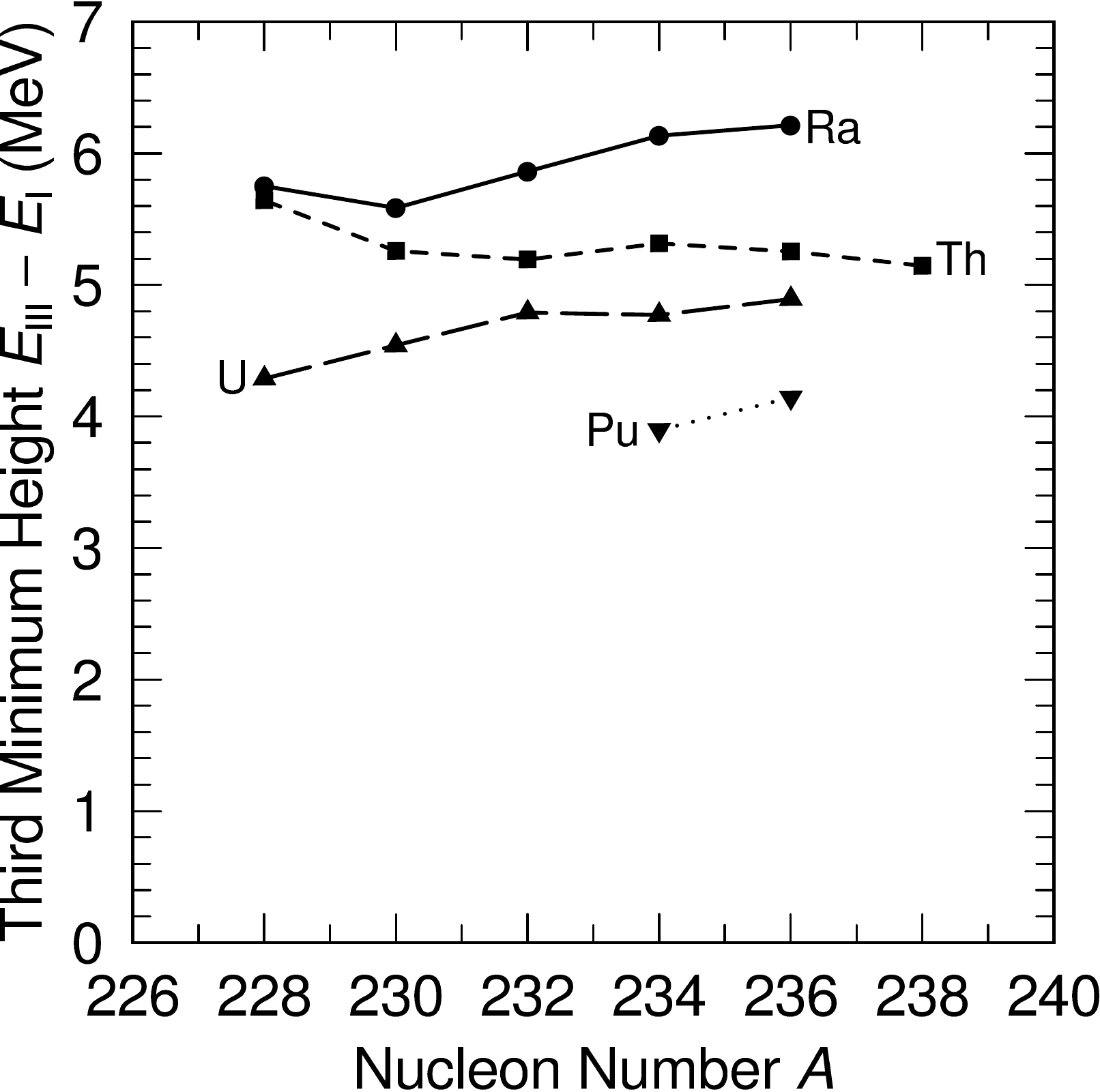} 
   \caption{Height of the third minimum with respect
to the ground state. 
The height of the third minimum decreases
 with \vspace{-0.3in}increasing $Z$.}
\label{thirhgt}
 \end{center}
\end{figure}
with spacings corresponding to moments of inertia consistent with
the large deformations of the calculated third minima.  More recent
experimental results were obtained for nearby nuclei by Csige and
collaborators \cite{csige09:a,csige12:a}.  For $^{232}$Pa the authors
obtained results similar to those obtained by Blons et al. for
nearby nuclei, namely a shallow, less than 1-MeV deep, third minimum at
about 5 MeV above the ground state.  However, their results for
$^{232}$U deviated from this pattern. They concluded the minimum was
surrounded by considerably higher peaks. According to their analysis
of the experimental data the third barrier peak lies 2.8 MeV above the
third minimum, which in turn is only 3.2 MeV above the ground state.

There are several  motivations to carry out yet another study of
third minima, despite the many previous studies.  First, the
results obtained in different models are sometimes mutually
inconsistent
\cite{moller72:a,moller74:a,moller74:b,bengtsson87:b,cwiok94:a,terakopian96:a,samyn05:a,moller09:a}
and we comment on possible reasons for these differences.  Second, we
have a more well-benchmarked model than those used earlier,
both in terms of the number of nuclides studied (globality)
and the number of properties calculated (universality). It is
therefore of interest to discuss results obtained within this
approach.

Constrained Hartree-Fock (and HFB) calculations, as applied so far, do not
unambiguously find the correct fission saddle points due to the inherent
limitations of constrained-minimization methods when applied to the
problem of locating saddle points, as is discussed in some detail in
Ref.\ \cite{moller09:a}, and is also pointed out by other authors
\cite{dubray12:a,kowal12:a}. All of the
earlier studies based on macroscopic-microscopic models, for example
\cite{moller72:a,bengtsson87:b}, investigated a very limited number of
shapes, less than 1000. Moreover, the multipole expansions used in many
of these studies are of questionable suitability for the elongated
shapes occurring in the later stages of fission.  Furthermore, the
modified-oscillator potential is very different from the folded-Yukawa
potential we use here.

Therefore, when we performed our first calculations in very large
deformation spaces \cite{moller99:a,moller00:a,moller01:a} we
anticipated a possible result could be that the third minima we had
seen in more limited calculations would no longer be present. However,
in these more detailed calculations we found, somewhat to our
surprise, similar third minima as found earlier.  In 
subsequent more extensive  large-scale
calculations of fission potential-energy surfaces we obtained results
for 5254 heavy nuclei \cite{moller09:a}.  For each nucleus the energy
was calculated for more than 5 million shapes as a function of five
deformation parameters: elongation or quadrupole moment, neck radius,
left nascent fragment spheroidal ($\epsilon_{\rm f1}$) deformation,
right nascent fragment spheroidal ($\epsilon_{\rm f2}$) deformation,
and nascent fragment mass asymmetry. We now analyze the calculated surfaces
specifically for third minima by immersion techniques (which were first
used in nuclear physics in Ref.~\cite{mamdouh98:a}) and all minima
deeper than 0.2 MeV and saddles between all possible pairwise
combinations of minima are determined and tabulated.  Full details of
the calculations are given in Ref.~\cite{moller09:a}. However, in
that paper third minima were discussed only briefly.

In Fig.~1 we show an example of a three-humped barrier, namely the
barrier of $^{236}$Ra.  This nucleus is currently not observed, but it
is interesting because it has one of the deeper third minima we have
found.  Also interesting is that at the second saddle peak,
the nascent fragments have not yet emerged. In contrast they are
already very prominent at the third peak.  This is also well expressed
in the increasingly negative value of the shell-plus-pairing
correction.  The shell-plus-pairing energy is -3.39 MeV, -1.724 MeV,
-5.63 MeV and -7.39 MeV at the fission-isomeric minimum, at the second
saddle, at the third minimum, and at the third (outermost) saddle,
respectively.  Some additional barriers of currently known nuclei, for
which we find third minima, are plotted in
Refs.~\cite{moller01:a,moller09:a}.  The reason the nascent fragments
have not yet emerged at the second peak is that this shape is too compact
to (geometrically) allow the formation of two distinct fragments with a
well-expressed narrow neck region in between.

In Fig.~2 we summarize our results for even-even systems.  Although we
have available the results for odd-even and odd-odd systems, we do not
include them here because they can be estimated by interpolation,
since barrier parameters change in a smooth manner with neutron and
proton number.  Fig.~2 shows the height of the second and third
barrier peaks relative to the third minimum.  We only show results for
nuclei where the third minimum is deeper than 0.2 MeV relative to the
lower of the two surrounding saddles. We only extend the plot to the
vicinity of the last known neutron-rich nuclei. Furthermore we exclude
many nuclei where we feel a third minimum would not be observable. For
example for $^{242}$U we find that the second peak is about 4 MeV
above the third minimum, whereas the third peak is only about 0.5 MeV
above. Therefore, because of the substantial second barrier peak,
resonances in the shallow third minimum would likely not be
observable.  A general result is that for the very lightest isotopes
we consider, the outer third peak is higher than the second peak. As
the neutron number increases the second peak becomes higher and the
third peak drops to lower energies and is also very low with
respect to the third minimum.  Therefore we find that the most likely
nuclei to exhibit experimental evidence of third minima in the barrier
are those shown in
Fig.~2 and nearby odd nuclei.  Possibly the best candidates for
observable third minima are those nuclei where the second and third
barrier peaks are of nearly equal height.  Thus our calculations
predict that $^{228}$Th--$^{236}$Th
and $^{228}$U--$^{236}$ U would exhibit the most clear experimental
signatures of third minima. The height of the third minimum
relative to the ground state is shown in Fig.~3.

Our results are very similar to the macroscopic-microscopic results shown
in Fig.~19 of
Ref. \cite{bengtsson87:b}, although a different potential and a limited
two-dimensional deformation space was used in that earlier calculation.
In contrast, the results of HFB \cite{samyn05:a} calculations are
highly variable. Quite different results are obtained for different
Skyrme forces, for example Samyn et al. \cite{samyn05:a} found
that BSk8 and SLy6$^{\delta\rho}$ give ``orthogonal'' results.  In
addition the outer saddle is often much too high relative to
experiment.  But we recall that many Skyrme forces have highly
non-optimum parameters and do not reproduce well other known
nuclear-structure data, such as nuclear masses. Furthermore, the
constrained-minimization method, by its very nature, can significantly
overestimate saddle-point energies \cite{moller09:a}. Therefore these
results may not yet be sufficiently realistic to predict reliably in
which nuclei third minima might be observable.

The results of {\v{C}}wiok et al. \cite{cwiok94:a} and Ter-Akopian
et al. \cite{terakopian96:a} show, in contrast to our results, 4-MeV deep
third minima for both light Th and U isotopes \cite{cwiok94:a} and the
heavy actinide $^{252}$Cf \cite{terakopian96:a}.  In their calculations
2D potential-energy surfaces were calculated as functions of quadrupole
and octupole moments $\beta_2$ and $\beta_3$, with the energy minimized
with respect to several additional, higher multipoles. This approach
suffers from the same mathematical limitations as the
``constrained'' self-consistent calculations discussed above. This has
now been confirmed in an independent study \cite{kowal12:a}. This
calculation, based on exactly the same macroscopic-microscopic
Woods-Saxon single-particle potential model that was used in the
{\v{C}}wiok et al. studies, employed instead immersion methods 
\cite{mamdouh98:a,moller01:a,moller09:a}, known to be
required to locate optimal saddle points on multi-dimensional
potential-energy surfaces. It was found, in close agreement
with our results using a different macroscopic-microscopic model, that
the third minima for Th are very shallow and only surrounded by
saddles less than 1 MeV high. Therefore, we feel it may be
unrealistic to expect such deep hyper-deformed minima to occur
in actinide nuclei.

Csige and collaborators \cite{csige09:a,csige12:a} have analyzed
experiments assuming a ``three-humped'' barrier. Somewhat in agreement
with the theoretical suggestion by {\v{C}}wiok et al. \cite{cwiok94:a} they
find for $^{232}$U a third minimum stabilized with respect to fission
by a 2.8 MeV high third peak relative to this third minimum. This
minimum is deduced to lie 3.2 MeV above the ground state. However for
the nearby nucleus $^{232}$Pa they find that the third minimum is less
than 1 MeV deep and lies at an excitation energy of 5 MeV above the
ground state. It is previously unheard of, both in experiments and
theoretical calculations, that fission-barrier parameters change by 
such large amounts between neighboring nuclei, differing by only one
unit in both $Z$ and $N$.  Therefore, we feel that in the analysis of
the experimental results, it is possible that choices made
for many of the parameters
in the cross-section model might have been  influenced by the
anticipation of barrier parameters similar to those obtained by
{\v{C}}wiok et al., which we now know \cite{kowal12:a} are incorrect.
Actual barrier curvatures and level-density parameters, needed in
cross-section models, may deviate
substantially from commonly used values. For example for the third
barrier peak in $^{232}$U we calculate a shell-plus-pairing correction
of $-4.85$ MeV, which translates to a very low level density here. At
the second peak the calculated shell-plus-pairing correction is -2.59
MeV, which implies a significantly different level density.

In conclusion, we have determined  the region of occurrence
of triple-humped fission
barriers. Only a relatively few nuclei in the light actinide region
exhibit this structure.  The isotope sequences $^{228}$Th--$^{236}$Th
and $^{228}$U--$^{236}$U (along with neighboring odd nuclei) appear
to be the best candidates for observable third minima.  Our results,
employing a folded-Yukawa single-particle potential in a 
macroscopic-microscopic model, are quite consistent both in terms of which
nuclei exhibit this feature and the heights of the saddle points and
minima with: 1) the experimental studies by Blons and collaborators
\cite{blons75:a,blons78:a,blons88:a}, 2) with early computer-limited
theoretical studies in both the oscillator and folded-Yukawa
macroscopic-microscopic models, and 3) with recent
state-of-the-art Woods-Saxon macroscopic-microscopic calculations
\cite{kowal12:a}
exploring a deformation space with about $1\,000\,000$ grid points.
 
This work was carried out under the auspices of the National
Nuclear Security Administration of the U. S. Department of
Energy at Los Alamos National Laboratory under Contract 
No.\ DE-AC52-06NA25396 and by travel grants for P. M.\ to JUSTIPEN
under grant number DE-FG02-06ER41407 (U. Tennessee)
Part of this research has been funded by MEXT HPCI STRATEGIC PROGRAM.

\end{document}